\documentclass[letterpaper, 10 pt, conference]{ieeeconf}  
\usepackage[dvipsnames]{xcolor}

\IEEEoverridecommandlockouts                              

\usepackage{amsmath}
\usepackage{amssymb}
\usepackage{amsfonts}

\usepackage{microtype}
\usepackage{graphicx}
\usepackage{subcaption}
\usepackage{booktabs} 

\usepackage[table]{xcolor}
\usepackage{graphicx}
\usepackage{afterpage}

\usepackage{graphicx} 

\usepackage{tikz}
\usepackage{pgfplots}
\pgfplotsset{compat=1.18}

\title{\LARGE \bf
Target Parameterization in Diffusion Models for Nonlinear Spatiotemporal System Identification
}

\author{Achraf El Messaoudi$^{1}$$^{*}$, Noureddine Khaous$^{2}$$^{*}$, Karim Cherifi$^{1}$
\thanks{$^{1}$ Université Marie et Louis Pasteur, SUPMICROTECH, CNRS, institut FEMTO-ST, F-25000 Besançon, France}
\thanks{$^{2}$ LLF, CNRS, Université Paris Cité, F-75013 Paris, France}
\thanks{$^{*}$ Equal contribution}
}

\begin{document}

\maketitle
\thispagestyle{empty}
\pagestyle{empty}

\begin{abstract}
Machine learning is becoming increasingly important for nonlinear system identification, including dynamical systems with spatially distributed outputs. However, classical identification and forecasting approaches become markedly less reliable in turbulent-flow regimes, where the dynamics are high-dimensional, strongly nonlinear, and highly sensitive to compounding rollout errors. Diffusion-based models have recently shown improved robustness in this setting and offer probabilistic inference capabilities, but many current implementations inherit target parameterizations from image generation, most commonly noise or velocity prediction. In this work, we revisit this design choice in the context of nonlinear spatiotemporal system identification. We consider a simple, self-contained patch-based transformer that operates directly on physical fields and use turbulent flow simulation as a representative testbed. Our results show that clean-state prediction consistently improves rollout stability and reduces long-horizon error relative to velocity- and noise-based objectives, with the advantage becoming more pronounced as the per-token dimensionality increases. These findings identify target parameterization as a key modeling choice in diffusion-based identification of nonlinear systems with spatial outputs in turbulent regimes.
\end{abstract}

\section{INTRODUCTION}
System identification for dynamical systems is a central problem in control, where the objective is to construct predictive models from measured data for simulation, analysis, and control design. In recent years, machine learning has become an increasingly important tool for this task, owing to its ability to capture complex dynamical behaviors directly from data \cite{pillonetto2025survey}.

Recent transformer-based approaches have started to broaden the scope of data-driven modeling for dynamical systems. In particular, sequence models have been used to infer the behavior of previously unseen systems directly from trajectory data \cite{forgione2023classmodels,rufolo2024enhanced, balim2023transformers}. Beyond these settings, related ideas have also been explored for parametric dynamical systems with time-varying inputs, where transformer architectures are used as black-box predictors for multi-output temporal evolution \cite{sun2025interpretable}.

These developments are also relevant for dynamical systems with spatial outputs, where the observations are no longer low-dimensional signals but distributed fields evolving over time. In this setting, the identification problem must account not only for temporal dynamics, but also for the spatial structure of the outputs. A natural extension of machine-learning-based state-space identification is therefore to preserve the nonlinear state-space viewpoint while replacing dense input-output mappings by convolutional parameterizations \cite{beintema2024thesis}.

For turbulent flows, however, this identification problem becomes substantially more difficult. The dynamics are high-dimensional, strongly nonlinear, and sensitive to compounding rollout errors, so that locally accurate predictors may still drift or become unstable in free-running simulation. In this regime, autoregressive conditional diffusion model (ACDM) has recently emerged as a promising approach, combining strong long-horizon prediction with posterior sampling capabilities that are attractive when multiple future flow evolutions are consistent with the same conditioning information \cite{kohl2023benchmarking}.

At the same time, several modeling choices in diffusion-based forecasting are inherited from image generation \cite{ho2020denoising,nichol2021improved,karras2022elucidating,peebles2022dit}. One important choice concerns the prediction objective: rather than predicting the clean next state directly, diffusion models are commonly trained to predict injected noise, or closely related parameterizations such as velocity \cite{ho2020denoising,karras2022elucidating}. However, recent work suggests that clean-signal prediction can be substantially easier than noise-like targets, especially in high-dimensional token spaces \cite{li2025back}. This observation naturally motivates revisiting the learning objective itself when identifying predictive surrogates for turbulent-flow dynamics.

Motivated by these considerations, we revisit prediction target choice in diffusion-based system identification for nonlinear distributed dynamics. Our goal is not to propose a highly specialized architecture, but rather to isolate how the diffusion prediction objective affects the quality of the identified model. To this end, we use a deliberately simple, self-contained patch-based transformer operating directly on physical fields, without latent autoencoders, auxiliary tokenizers, or external backbone components, and vary only the diffusion target while keeping the architecture and training budget fixed. Concretely, we compare clean-state, velocity, and noise prediction on the incompressible and transonic benchmarks introduced in \cite{kohl2023benchmarking}, with an emphasis on long-horizon rollouts. 
These observations suggest that, in diffusion-based system identification of nonlinear distributed dynamics, the prediction target should be regarded as a genuine modeling choice rather than a secondary implementation detail. This motivates the controlled study carried out in this paper, where we isolate the effect of target parameterization on the quality of the identified transition model.

The contributions of this paper are as follows:
\begin{itemize}
    \item We revisit the choice of diffusion prediction objective ($x$-, $v$-, or $\epsilon$-prediction) in diffusion-based system identification of turbulent-flow dynamics, and provide a controlled evaluation on incompressible and transonic cylinder benchmarks under matched training budgets.
    \item We introduce a minimal, self-contained patch-transformer diffusion system identification model for next-step field prediction that operates directly in physical space and conditions on past frames and scalar parameters, enabling a clean assessment of objective-choice effects without encoders, U-Nets, or latent variables.
    \item We show that clean-state prediction leads to more reliable identified transition models than conventional noise-like targets, yielding improved rollout stability and lower long-horizon error, and we relate this behavior to a manifold-based interpretation of physical states versus diffusion targets.
    \item We propose a two-resolution, matched-token-count protocol that varies per-token dimensionality while keeping sequence length fixed, and show that the advantage of clean-state prediction becomes more pronounced in the large-patch regime.
\end{itemize}

\section{BACKGROUND AND PROBLEM SETUP}
\label{sec:background}

We consider dynamical systems with spatial outputs, for which collected input-output trajectory data are available. The spatial outputs are spatio-temporal fields $\{x_t\}_{t=1}^{T}$ with
$x_t \in \mathbb{R}^{C \times H \times W}$, where $C$ denotes the number of physical channels and $(H,W)$ the spatial resolution.
Here, $T$ denotes the number of time steps in a trajectory.
In many control-oriented settings, each trajectory is associated with two types of exogenous information: a time-varying input signal $u_{1:T}$ driving the system evolution, and possibly a fixed low-dimensional parameter vector $\theta$ describing the operating condition or system instance.
Accordingly, from a system-identification viewpoint, the objective is to identify a predictive model of the field evolution from past observations together with the available control inputs and, when relevant, static parameters. This setting arises naturally in nonlinear systems with spatially distributed outputs, where one seeks to model the evolution of measured fields under external actuation. Representative examples are discussed in \cite{beintema2024thesis}, including flow around an oscillating cylinder, where the transverse cylinder displacement acts as the input, and Rayleigh--B\'enard convection, where the manipulated input is the temperature difference imposed between two bottom wall sections.

We model the conditional next-step distribution given a fixed-length history window together with the available exogenous information:
\begin{equation}
\hat{x}_{t+1} \sim p_\phi(\,\cdot \mid x_{t-k+1:t},\, u_{t-k+1:t},\, \theta),
\label{eq:ar_conditional}
\end{equation}
where $k$ denotes the context length, $x_{t-k+1:t}:=(x_{t-k+1},\ldots,x_t)$, and $u_{t-k+1:t}:=(u_{t-k+1},\ldots,u_t)$.

A free-running rollout of horizon $H$ is then obtained by recursively replacing the true past states in \eqref{eq:ar_conditional} with previously generated ones, for $t = k,\ldots,k+H-1$.
\label{eq:rollout}
Here, $\phi$ denotes the learnable parameters of the conditional generative model $p_\phi$.
From a system-identification perspective, this corresponds to evaluating the identified surrogate in free-running simulation, which is the regime relevant for long-horizon prediction and downstream control-oriented use.
A central difficulty in this setting is compounding error: small inaccuracies at each step can accumulate during rollout and lead to drift or instability, a phenomenon widely observed in autoregressive sequence prediction and particularly severe in chaotic flow dynamics.

\begin{figure}[t]
  \centering
  \includegraphics[width=1.2\linewidth]{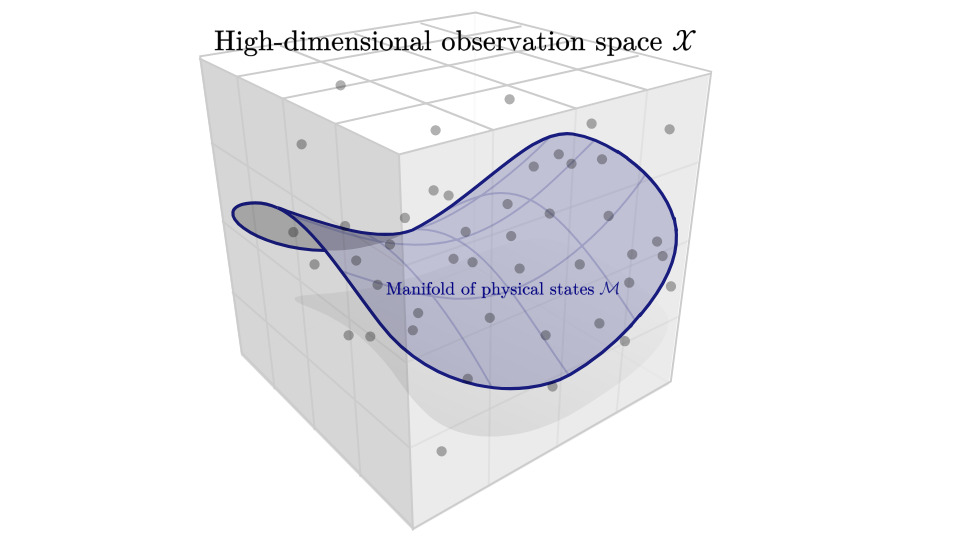}
  \caption{\textbf{Geometric assumption.}
  Under the manifold assumption, physical states $x \in \mathcal{X}$ concentrate near a low-dimensional subset $\mathcal{M}\subset \mathcal{X}$, while injected noise $\epsilon \sim \mathcal{N}(0,I)$ is isotropic in $\mathcal{X}$.}
  \label{fig:manifold}
\end{figure}

\subsection{Conditional diffusion / rectified flow for next-step generation}
\label{sec:rf_setup}
We instantiate \eqref{eq:ar_conditional} using a diffusion-style generative model, adopting a rectified-flow formulation that learns a velocity field transporting noise to data.
To avoid confusion with the discrete time index $t$ in $\{x_t\}$, we denote the continuous diffusion time by $\tau\in[0,1]$.
Given a target next state $x \equiv x_{t+1}$ and conditioning information $c := (x_{t-k+1:t}, \theta)$, we sample $\tau \in [0,1]$ and form a noised interpolation
\begin{equation}
z_\tau = \tau\,x + (1-\tau)\,\epsilon,
\qquad \epsilon \sim \mathcal{N}(0, I).
\label{eq:zt}
\end{equation}

For the linear coupling in \eqref{eq:zt}, the induced \emph{target} velocity along the path is
\begin{equation}
v^\star(z_\tau,\tau;x,\epsilon)
\;:=\;
\frac{\mathrm{d}z_\tau}{\mathrm{d}\tau}
\;=\;
x-\epsilon
\;=\;
\frac{x-z_\tau}{1-\tau},
\label{eq:vstar}
\end{equation}
where $v^\star$ denotes the ground-truth velocity implied by the coupling, while $v_\phi(z,\tau,c)$ denotes the learned velocity field parameterized by $\phi$.

At inference time, we generate $\hat{x}$ by integrating the learned ODE from $\tau=0$ (noise) to $\tau\approx 1$ using a standard numerical solver : 
\begin{equation}
\frac{\mathrm{d}z}{\mathrm{d}\tau} = v_\phi(z,\tau,c),\quad z(0)\sim \mathcal{N}(0,I), \quad \hat{x}=z(1-\varepsilon),
\label{eq:ode}
\end{equation}
with a small $\varepsilon>0$ for numerical stability.
This continuous-time view is closely connected to diffusion/score-based modeling \cite{ho2020denoising} and provides a convenient interface for conditional next-step generation within the autoregressive rollout \eqref{eq:rollout}.

\subsection{Prediction target parameterizations}
\label{sec:targets}
A central modeling choice is the training target predicted from $(z_\tau,\tau,c)$.
We consider three standard parameterizations widely used in diffusion-style models:
\begin{itemize}
\item \textbf{$x$-prediction:} predicts the clean next state $\hat{x}_\phi(z_\tau,\tau,c)$.
\item \textbf{$\epsilon$-prediction:} predicts the injected noise $\hat{\epsilon}_\phi(z_\tau,\tau,c)$.
\item \textbf{$v$-prediction:} predicts the velocity $\hat{v}_\phi(z_\tau,\tau,c)$ as in \eqref{eq:vstar}.
\end{itemize}
Under \eqref{eq:zt}, these targets are analytically related; for example,
$x = \frac{z_\tau - (1-\tau)\epsilon}{\tau}$ (for $\tau>0$) and $v = x - \epsilon$.
Accordingly, in idealized settings the parameterizations can be transformed into one another and are often treated as interchangeable up to reweighting \cite{ho2020denoising,nichol2021improved,karras2022elucidating}.

A different conclusion emerges once one accounts for finite capacity and representation bottlenecks.
A common hypothesis in high-dimensional learning is that natural signals concentrate near a low-dimensional manifold of the ambient space.
Denoising can then be viewed as recovering a structured signal while suppressing components orthogonal to this subset, which is an interpretation that also underlies denoising autoencoders. In contrast, Gaussian noise $\epsilon \sim \mathcal{N}(0,I)$ is isotropic in the full ambient space.
This geometric asymmetry motivates why learning to reconstruct $x$ can differ from learning to predict noise-like quantities (such as $\epsilon$, or the noise component implicit in $v=x-\epsilon$) despite analytic equivalences in \eqref{eq:zt}; Figure~\ref{fig:manifold} summarizes the intuition.

This perspective is emphasized in \cite{li2025back}, where the authors introduce JiT (\emph{Just Image Transformer}) and argue that the practical equivalence between $x$- and $\epsilon$-prediction can break sharply for transformer diffusion models when the representation is under-complete relative to the target space. More concretely, patch tokenization introduces an explicit per-token ambient dimension: with patch size $P$, each token corresponds to a vector in $\mathbb{R}^{CP^2}$, and this dimension grows quadratically with $P$. When $CP^2$ is large relative to the model width, predicting noise-like targets can become substantially more demanding than predicting structured signals, even though the targets are analytically related under \eqref{eq:zt}. In the remainder of this paper, we refer to our autoregressive adaptation of this architecture as A-JiT (\emph{Autoregressive JiT}).

\section{Method}
\label{sec:method}

\begin{figure}[t]
  \centering
  \includegraphics[width=1.0\linewidth]{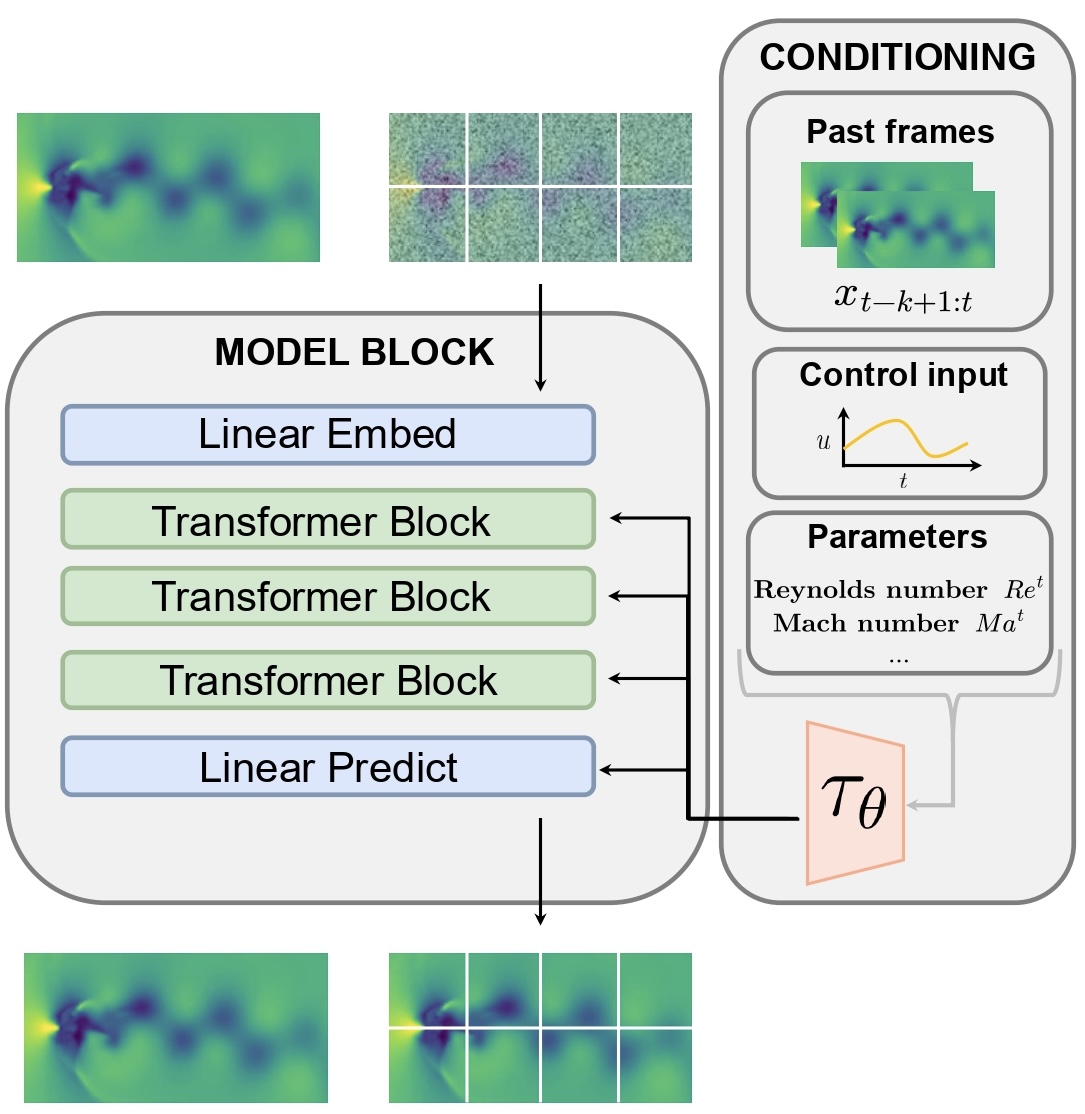}
  \caption{\textbf{Method overview.} A self-contained patch-token transformer maps the noised next-state $z_\tau$ to a field-level prediction, conditioned on the history window $x_{t-k+1:t}$, diffusion time, and parameters $\theta$.}
  \label{fig:arch_overview}
\end{figure}

We instantiate the autoregressive next-step model $p_\phi(\cdot\mid x_{t-k+1:t},u_{t-k+1:t},\theta)$ using a conditional rectified-flow formulation (Section~\ref{sec:rf_setup}) parameterized by a patch-token transformer backbone (Figure~\ref{fig:arch_overview}).
The architecture is deliberately self-contained: it operates directly in physical field space via patch tokens (no latent encoder/decoder), so that differences in long-horizon rollouts can be attributed primarily to the diffusion target parameterization rather than auxiliary architectural components.
We use the notation from Section~\ref{sec:background}: the next state is $x \equiv x_{t+1}$, the conditioning is $c := (x_{t-k+1:t},u_{t-k+1:t},\theta)$, and the noised input is $z_\tau$ in \eqref{eq:zt}.
Across all experiments, we keep the backbone fixed and vary only the training target ($x$, $v$, or $\epsilon$).

\subsection{Self-contained patch-token transformer backbone}
\label{sec:arch}

Figure~\ref{fig:arch_overview} summarizes the forward path. The ``Pixel/Field space'' blocks correspond to the noised input $z_\tau$ and the predicted output in $\mathbb{R}^{C\times H\times W}$. The ``Model block'' consists of patchification with a linear embedding into tokens, a stack of transformer blocks, and a linear head followed by unpatchification. Conditioning enters through two complementary routes: a token-level encoding of the past window $x_{t-k+1:t}$ (added to the input tokens), and a global pathway that injects $(\tau,\theta)$ into each transformer block via feature-wise modulation.

\paragraph{Patch tokenization}
Let $P$ be the patch size (dividing $H$ and $W$). A field $u\in\mathbb{R}^{C\times H\times W}$ is partitioned into
$N=\frac{H}{P}\cdot\frac{W}{P}$ non-overlapping patches and each flattened patch in $\mathbb{R}^{CP^2}$ is mapped to a token in $\mathbb{R}^{D}$.
In practice, we implement this with a strided convolution (kernel size and stride equal to $P$), which is equivalent to a shared linear projection across patches.
We denote patch embedding by $\mathrm{Tok}(\cdot)$, yielding $\mathrm{Tok}(u)\in\mathbb{R}^{N\times D}$.

\paragraph{History conditioning (token-level)}
We stack the $k$ context frames along channels,
$x_{t-k+1:t}^{\mathrm{stk}}\in\mathbb{R}^{(kC)\times H\times W}$, and embed them into tokens
$T_{\mathrm{hist}}=\mathrm{Tok}_{\mathrm{hist}}(x_{t-k+1:t}^{\mathrm{stk}})\in\mathbb{R}^{N\times D}$.
Similarly, we embed the diffusion input into $T_z=\mathrm{Tok}_z(z_\tau)\in\mathbb{R}^{N\times D}$.
The token sequence entering the transformer is the token-wise sum
\begin{equation}
T_0 \;=\; T_z \;+\; T_{\mathrm{hist}} \;+\; E_{\mathrm{pos}},
\label{eq:token_sum}
\end{equation}
where $E_{\mathrm{pos}}\in\mathbb{R}^{N\times D}$ is a fixed 2D sinusoidal positional embedding.

This design keeps the attention sequence length equal to the patch-grid size and enables token-wise fusion of the current diffusion state $z_\tau$ with the conditioning history.

\paragraph{Global conditioning via $(\tau,\theta)$}
We embed diffusion time $\tau$ using sinusoidal/Fourier features followed by an MLP (as commonly used in diffusion models), and embed the parameter vector $\theta$ with an MLP. Their sum gives a global conditioning vector
\begin{equation}
g(\tau,\theta)=g_\tau(\tau)+g_\theta(\theta)\in\mathbb{R}^{D},
\label{eq:global_cond}
\end{equation}
which modulates each transformer block through feature-wise shift/scale (FiLM/AdaLN-style) \cite{peebles2022dit}.
Intuitively, this pathway provides per-step context (diffusion time) and trajectory identity (e.g., Re/Ma) without increasing the token sequence length.

\paragraph{Transformer and field output}
We apply a stack of $L$ transformer blocks with multi-head self-attention and MLP sublayers, following the ViT/DiT pattern.
Let $T_L\in\mathbb{R}^{N\times D}$ denote the final token sequence.
A linear head maps each token back to a $\mathbb{R}^{CP^2}$ patch vector, and unpatchification reconstructs a field output
\begin{equation}
y_\phi(z_\tau,\tau,c)\in\mathbb{R}^{C\times H\times W}.
\label{eq:model_output}
\end{equation}
The semantic meaning of $y_\phi$ depends on the target parameterization below.

\subsection{Targets and training objectives}
\label{sec:method_targets}

All variants share the same backbone (Section~\ref{sec:arch}); only the interpretation of $y_\phi$ and the regression objective differ. For each training example, we sample $\tau\in[0,1]$, draw $\epsilon\sim\mathcal{N}(0,I)$, and construct $z_\tau$ via \eqref{eq:zt}.

\paragraph{$x$-prediction.}
We set $\hat{x}_\phi = y_\phi(z_\tau,\tau,c)$ and match the induced velocity along the coupling path :
\begin{equation}
\hat{v}_\phi \;=\; \frac{\hat{x}_\phi - z_\tau}{1-\tau},
\label{eq:v_from_xhat}
\end{equation}
regressing to $v^\star=\frac{x-z_\tau}{1-\tau}$:
\begin{equation}
\mathcal{L}_{x\rightarrow v}(\phi)
=
\mathbb{E}\Big[\ell\big(\hat{v}_\phi,\, v^\star\big)\Big].
\label{eq:loss_x2v}
\end{equation}

\paragraph{$v$-prediction.}
We set $\hat{v}_\phi = y_\phi(z_\tau,\tau,c)$ and minimize
\begin{equation}
\mathcal{L}_{v}(\phi)
=
\mathbb{E}\Big[\ell\big(\hat{v}_\phi,\, v^\star\big)\Big].
\label{eq:loss_v}
\end{equation}

\paragraph{$\epsilon$-prediction.}
We set $\hat{\epsilon}_\phi = y_\phi(z_\tau,\tau,c)$ and minimize
\begin{equation}
\mathcal{L}_{\epsilon}(\phi)
=
\mathbb{E}\Big[\ell\big(\hat{\epsilon}_\phi,\, \epsilon\big)\Big],
\label{eq:loss_eps}
\end{equation}
as in DDPM-style training \cite{ho2020denoising}.

\subsection{Sampling and autoregressive rollout}
\label{sec:sampling_rollout}

Sampling integrates the ODE in \eqref{eq:ode} and therefore requires a velocity field $v_\phi(z,\tau,c)$.
For $v$-prediction we set $v_\phi\equiv\hat{v}_\phi$.
For $x$-prediction we compute $v_\phi$ from $\hat{x}_\phi$ using \eqref{eq:v_from_xhat}.
For $\epsilon$-prediction we first recover $\hat{x}_\phi = \frac{z_\tau-(1-\tau)\hat{\epsilon}_\phi}{\tau}$ (for $\tau>0$) and then apply \eqref{eq:v_from_xhat}.
We integrate up to $\tau=1-\varepsilon$ for numerical stability, following common practice in continuous-time diffusion/flow samplers.

\paragraph{ODE discretization.}
We solve the rectified-flow ODE on $\tau\in[0,1-\varepsilon]$ using a fixed-step explicit integrator.
Unless stated otherwise, we use Heun's method (predictor--corrector), and report Euler as a simpler alternative. 
Let $\tau_0=0<\cdots<\tau_M=1-\varepsilon$ with step size $\Delta\tau=\tau_{i+1}-\tau_i$.
Starting from $z_0\sim\mathcal{N}(0,I)$, the update uses the learned velocity field $v_\phi(z,\tau,c)$ and returns $\hat{x}=z_M$.

Given a history window, control inputs, and parameters, we sample $\hat{x}_{t+1}\sim p_\phi(\cdot\mid \hat{x}_{t-k+1:t},u_{t-k+1:t},\theta)$ and feed it back into the context window, iterating for the desired horizon as in \eqref{eq:rollout}.
This free-running protocol matches the standard evaluation used in autoregressive diffusion turbulence benchmarks.

\section{Experiments}
\label{sec:experiments}

\subsection{Datasets}
\label{sec:datasets}

We evaluate our diffusion-based system identification method on two 2D cylinder-flow benchmarks released with ACDM \cite{kohl2023benchmarking}: an incompressible vortex-shedding wake (\texttt{Inc}) and a compressible transonic cylinder flow (\texttt{Tra}). Both datasets provide spatio-temporal fields on a regular $128\times 64$ grid along with a scalar parameter $\theta$ (Reynolds or Mach number). The proposed framework can be extended to settings where the field evolution is conditioned not only on past frames but also on time-varying control inputs. In the present work, however, we restrict attention to these parameter-conditioned benchmarks, since they provide a standardized testbed for direct comparison with ACDM. Following the original protocol, we use standardized preprocessing, per-channel normalization, and obstacle masking for both training and evaluation.

\paragraph{\texttt{Inc}: incompressible cylinder wake.}
This benchmark consists of periodic vortex shedding across Reynolds numbers $\mathrm{Re}\in[100,1000]$. The fields include velocity and pressure channels, with $\theta=\mathrm{Re}$ serving as the scalar conditioning parameter. 

\paragraph{\texttt{Tra}: transonic cylinder flow.}
\texttt{Tra} contains compressible flows at a fixed Reynolds number while varying the freestream Mach number $\mathrm{Ma}\in[0.5,0.9]$. The data includes velocity, density, and pressure fields, with $\theta=\mathrm{Ma}$ for conditioning. We report metrics on the specific subset of modeled channels as detailed in our experiments.

\paragraph{Splits and horizons.}
Following \cite{kohl2023benchmarking}, we adopt evaluation regimes that probe generalization in $\theta$ (interpolation and extrapolation) as well as temporal stability through extended rollouts. Exact train/test splits, horizons, and channel configurations are specified in Sec. \ref{sec:exp_setup}.

\subsection{Experimental protocol}
\label{sec:exp_setup}

Our experiments isolate the effect of the diffusion \emph{target parameterization} ($x$-, $v$-, or $\epsilon$-prediction; Section~\ref{sec:method_targets}) on \emph{autoregressive rollout quality} in a system identification setting, while keeping the backbone, optimizer, and sampling procedure fixed.
Models are trained for one-step conditional generation and evaluated in free-running rollouts where predictions are recursively fed back as inputs.

\paragraph{Two-resolution protocol with matched token count.}
A central diagnostic is a controlled change in the \emph{per-token ambient dimension} induced by patch tokenization.
With patch size $P$, each token corresponds to a vector in $\mathbb{R}^{C P^2}$ (Section~\ref{sec:arch}), and the number of tokens is $N=(H/P)(W/P)$.
We compare two spatial resolutions while keeping $N$ fixed:
\begin{itemize}
\item \textbf{Low resolution:} $H\times W = 64\times 32$ with patch size $P=4$,
\item \textbf{High resolution:} $H\times W = 256\times 128$ with patch size $P=16$.
\end{itemize}
In both cases, $N=(H/P)(W/P)=16\cdot 8=128$ tokens, so self-attention operates over the same sequence length.
However, the per-token ambient dimension $C P^2$ increases by a factor of $16$ from $P=4$ to $P=16$.
This separates effects due to token \emph{count} from effects due to token \emph{dimension}, which is precisely the regime where target choice is expected to matter most for patch-token diffusion transformers.

\paragraph{Target-choice variants.}
For each dataset (\texttt{Inc}, \texttt{Tra}) and each resolution setting above, we train three models that share the \emph{same} backbone and training hyperparameters, differing only in the predicted target and loss:
\begin{enumerate}
\item \textbf{$x$-prediction with induced velocity loss:} the network outputs $\hat{x}_\phi$ and is trained via $\mathcal{L}_{x\rightarrow v}$ in \eqref{eq:loss_x2v};
\item \textbf{$v$-prediction:} the network outputs $\hat{v}_\phi$ and is trained with $\mathcal{L}_v$ in \eqref{eq:loss_v};
\item \textbf{$\epsilon$-prediction:} the network outputs $\hat{\epsilon}_\phi$ and is trained with $\mathcal{L}_{\epsilon}$ in \eqref{eq:loss_eps}.
\end{enumerate}
This matched design isolates the practical impact of target parameterization under fixed capacity and tokenization.

\paragraph{Training, sampling, and compute matching.}
All variants are trained on next-step examples built from trajectories using context length $k=2$.
At inference time, each next-step prediction is obtained by integrating the rectified-flow ODE \eqref{eq:ode} up to $\tau=1-\varepsilon$ using a fixed-step solver, and then rolled out autoregressively for a fixed horizon.
To ensure fair comparisons, we keep the backbone architecture (depth, width, heads), batch size, learning-rate schedule, and the total number of training updates fixed across targets and across the two-resolution protocol.
Since $N$ is matched between the two resolutions, the attention sequence length is comparable; the primary controlled change is the per-token ambient dimension $C P^2$.

\subsection{Evaluation metrics}
\label{sec:metrics}

We evaluate autoregressive simulators along three complementary axes: pointwise accuracy, temporal stability, and temporal spectral fidelity.
All metrics are computed on free-running rollouts \eqref{eq:rollout} and aggregated over rollout time and test trajectories.
For cylinder-flow data, we exclude the obstacle region using the provided mask whenever it is available and applicable.

\paragraph{Pointwise accuracy}
We report the mean-squared error (MSE) between predicted and reference fields outside the obstacle region.
While MSE can be sensitive to phase shifts in oscillatory regimes, it provides a standardized baseline for comparing rollouts under a fixed protocol.

\paragraph{Temporal stability}
For each test simulation indexed by $q\in\mathcal{Q}$, we generate a free-running rollout
$\{\hat{x}^{(q,s)}_t\}_{t=0}^{T-1}$ of length $T$, where $s\in\{1,\dots,S\}$ indexes independent stochastic samples
(i.e., different diffusion sampling noise with the same conditioning).
Let $\Delta t$ denote the time increment between consecutive frames and let
$\hat{x}^{(q,s)}_t\in\mathbb{R}^{C\times H\times W}$ be the predicted field at rollout step $t$.

We measure the mean absolute time derivative between consecutive predicted frames:
\begin{equation}
d^{(q,s)}_t
\;=\;
\frac{1}{CHW}\sum_{c=1}^{C}\sum_{i=1}^{H}\sum_{j=1}^{W}
\frac{\hat{x}^{(q,s)}_{t}(c,i,j)-\hat{x}^{(q,s)}_{t-1}(c,i,j)}{\Delta t}
\label{eq:temporal_change}
\end{equation}

We summarize temporal stability as a function of rollout step by averaging across all test simulations and stochastic samples:
\begin{equation}
\bar d_t
\;=\;
\frac{1}{|\mathcal{Q}|S}\sum_{q\in\mathcal{Q}}\sum_{s=1}^{S} d^{(q,s)}_t,
\qquad t=1,\dots,T-1,
\label{eq:temporal_change_mean}
\end{equation}
and we visualize dispersion with a min--max envelope over the same set of $(q,s)$.

For the ground-truth rollout $\{x^{(q)}_t\}_{t=0}^{T-1}$, we compute the analogous quantity
$d^{(q)}_t$ by replacing $\hat{x}$ with $x$, and aggregate across $q\in\mathcal{Q}$ to obtain the reference curve.
Large values of $\bar d_t$ indicate temporally noisy rollouts or drift, whereas abnormally small values can indicate over-damping
and collapse toward a nearly steady state.

\paragraph{Spectral fidelity}
To probe whether long rollouts preserve the dominant temporal dynamics, we extract a scalar time signal from a fixed spatial location.
Let $p=(i_p,j_p)$ denote the probe point and let $c^\star$ be the channel of interest (here we use the modeled field channel at $p$).
For each test simulation $q\in\mathcal{Q}$ and stochastic sample $s\in\{1,\dots,S\}$, we define the probe signal
\begin{equation}
s^{(q,s)}_t \;=\; \hat{x}^{(q,s)}_t(c^\star,i_p,j_p),
\qquad t=0,\dots,T-1 .
\label{eq:probe_signal}
\end{equation}
We compute its discrete Fourier transform
\begin{equation}
\widehat{s}^{(q,s)}(f_m)
\;=\;
\sum_{t=0}^{T-1} s^{(q,s)}_t \, e^{-i2\pi f_m t \Delta t},
\qquad
f_m=\frac{m}{T\Delta t},
\label{eq:probe_dft}
\end{equation}
and report the one-sided power spectrum
\begin{equation}
P^{(q,s)}(f_m)
\;=\;
\big|\widehat{s}^{(q,s)}(f_m)\big|^2,
\qquad m=1,\dots,\left\lfloor\tfrac{T}{2}\right\rfloor-1 .
\label{eq:probe_psd}
\end{equation}
In figures, we follow the plotting convention used in our analysis code and visualize the frequency-weighted spectrum
$f_m^{2}\,P^{(q,s)}(f_m)$ to highlight discrepancies at higher temporal frequencies.

We average $f_m^{2}\,P^{(q,s)}(f_m)$ across all $(q,s)$ pairs to obtain the reported spectrum and visualize dispersion using a min--max envelope.
For the reference simulation, we repeat the same procedure using $x^{(q)}_t$ instead of $\hat{x}^{(q,s)}_t$.

\section{Experimental Results}
\label{sec:results}

\subsection{Target choice}
\label{sec:results_target_choice}

We study how the choice of prediction target and regression objective affects autoregressive rollouts under a fixed backbone and training budget. Concretely, we evaluate a $3\times 3$ design over targets ($x$, $\epsilon$, $v$) and regression objectives ($x$-, $\epsilon$-, $v$-loss). Tables~\ref{fig:inc_targetloss} and \ref{fig:tra_targetloss} report the resulting long-horizon rollout error for \texttt{INC} and \texttt{TRA} under the two-resolution, matched-token-count protocol ($N=128$).

Overall, $x$-prediction yields the most reliable rollouts. Crucially, the gap with $\epsilon$- and $v$-based parameterizations widens in the high-resolution setting, where the patch size is larger ($P{=}16$) and each token corresponds to a higher-dimensional patch vector in $\mathbb{R}^{C P^2}$. This regime increases the difficulty of regressing noise-like targets at fixed model width, consistent with the token-dimension perspective discussed in Section~\ref{sec:targets}. From a compute standpoint, moving from $(64{\times}32,P{=}4)$ to $(256{\times}128,P{=}16)$ keeps the token count fixed but increases the cost of the input/output patch projections (which scale with $CP^2$) and the reconstruction and loss evaluation on a larger grid, resulting in noticeably longer training time per step.

\begin{table}[t]
\centering
\small
\setlength{\tabcolsep}{10pt}
\renewcommand{\arraystretch}{1.25}

\begin{tabular}{l|c|c|c}
\toprule
 & $\boldsymbol{x}$-pred & $\boldsymbol{\epsilon}$-pred & $\boldsymbol{v}$-pred \\
\midrule
$\boldsymbol{x}$-loss &
\cellcolor{green!15}0.021 &
\cellcolor{red!10}10.32 &
\cellcolor{red!10}14.12 \\
$\boldsymbol{\epsilon}$-loss &
\cellcolor{green!15}0.028 &
\cellcolor{red!10}10.41 &
\cellcolor{red!10}15.09 \\
$\boldsymbol{v}$-loss &
\cellcolor{green!15}0.018 &
\cellcolor{red!10}9.28 &
\cellcolor{red!10}13.10 \\
\bottomrule
\end{tabular}

\vspace{4pt}
{\footnotesize (a) \texttt{INC} $256\times128$, $P{=}16$}

\vspace{10pt}

\begin{tabular}{l|c|c|c}
\toprule
 & $\boldsymbol{x}$-pred & $\boldsymbol{\epsilon}$-pred & $\boldsymbol{v}$-pred \\
\midrule
$\boldsymbol{x}$-loss &
\cellcolor{green!15}0.011 &
\cellcolor{green!15}0.017 &
\cellcolor{green!15}0.016 \\
$\boldsymbol{\epsilon}$-loss &
\cellcolor{green!15}0.012 &
\cellcolor{green!15}0.010 &
\cellcolor{green!15}0.013 \\
$\boldsymbol{v}$-loss &
\cellcolor{green!15}0.009 &
\cellcolor{green!15}0.012 &
\cellcolor{green!15}0.01 \\
\bottomrule
\end{tabular}

\vspace{4pt}
{\footnotesize (b) \texttt{INC} $64\times32$, $P{=}4$}

\caption{Target and loss choices on benchmark \texttt{INC}.}
\label{fig:inc_targetloss}
\end{table}

\begin{table}[t]
\centering
\small
\setlength{\tabcolsep}{10pt}
\renewcommand{\arraystretch}{1.25}

\begin{tabular}{l|c|c|c}
\toprule
 & $\boldsymbol{x}$-pred & $\boldsymbol{\epsilon}$-pred & $\boldsymbol{v}$-pred \\
\midrule
$\boldsymbol{x}$-loss &
\cellcolor{green!15}0.026 &
\cellcolor{red!10}12.27 &
\cellcolor{red!10}11.12 \\
$\boldsymbol{\epsilon}$-loss &
\cellcolor{green!15}0.029 &
\cellcolor{red!10}13.36 &
\cellcolor{red!10}14.13 \\
$\boldsymbol{v}$-loss &
\cellcolor{green!15}0.022 &
\cellcolor{red!10}11.22 &
\cellcolor{red!10}12.10 \\
\bottomrule
\end{tabular}

\vspace{4pt}
{\footnotesize (a) \texttt{TRA} $256\times128$, $P{=}16$}


\begin{tabular}{l|c|c|c}
\toprule
 & $\boldsymbol{x}$-pred & $\boldsymbol{\epsilon}$-pred & $\boldsymbol{v}$-pred \\
\midrule
$\boldsymbol{x}$-loss &
\cellcolor{green!15}0.014 &
\cellcolor{green!15}0.018 &
\cellcolor{green!15}0.016 \\
$\boldsymbol{\epsilon}$-loss &
\cellcolor{green!15}0.015 &
\cellcolor{green!15}0.016 &
\cellcolor{green!15}0.020 \\
$\boldsymbol{v}$-loss &
\cellcolor{green!15}0.011 &
\cellcolor{green!15}0.014 &
\cellcolor{green!15}0.015 \\
\bottomrule
\end{tabular}

\vspace{4pt}
{\footnotesize (b) \texttt{TRA} $64\times32$, $P{=}4$}

\caption{Target and loss choices on benchmark \texttt{TRA}.}
\label{fig:tra_targetloss}
\end{table}

\subsection{Ablation Study: Bottleneck Linear Embedding}

To further investigate the manifold assumption in the context of chaotic PDE simulations, we analyze the effect of a \textbf{bottleneck linear embedding} on rollout stability. Following the findings in \cite{kohl2023benchmarking}, we replace the standard linear patch projection $Tok(u) \in \mathbb{R}^{N \times D}$ with a pair of sequential linear layers characterized by an intermediate bottleneck dimension $d'$, where $d' < \min(CP^2, D)$. Formally, the embedding of a flattened physical patch $p \in \mathbb{R}^{CP^2}$ is redefined as:
\begin{equation}
    E_{patch} = W_2 (W_1 p + b_1) + b_2
\end{equation}
where $W_1 \in \mathbb{R}^{d' \times CP^2}$ and $W_2 \in \mathbb{R}^{D \times d'}$. This structure acts as a \textbf{low-rank reparameterization} of the input space.

\begin{figure}[t]
    \centering
    \begin{tikzpicture}
        \begin{axis}[
            width=\columnwidth,
            height=5cm,
            title={\small \textbf{Global MSE Sum vs. Dimension $d'$}},
            xlabel={\footnotesize $d'$ (Dimension)},
            ylabel={\footnotesize Total MSE $(\times 10^{-2})$}, 
            xmode=log,
            log basis x={2},
            xtick={2, 4, 8, 16, 32, 64},
            xticklabels={2, 4, 8, 16, 32, 64},
            grid=both,
            grid style={line width=.1pt, draw=gray!10},
            major grid style={line width=.2pt, draw=gray!30},
            tick label style={font=\tiny},
            title style={font=\footnotesize},
            label style={font=\footnotesize},
            legend style={font=\tiny, at={(0.05,0.95)}, anchor=north west}
        ]
            \addplot[
                red, 
                thick, 
                dashed, 
                domain=2:64, 
                samples=2
            ] {2.5}; 
            \addlegendentry{No Bottleneck}

            \addplot[
                color=blue,
                mark=*,
                thick,
                mark options={fill=blue, scale=0.8}
            ] coordinates {
                (2, 2.2)    
                (4, 0.98)   
                (8, 1.25)   
                (16, 1.4)   
                (32, 2.4)   
                (64, 4.6)   
            };
            \addlegendentry{Bottleneck}

        \end{axis}
    \end{tikzpicture}

    \caption{Sum of Global MSE across dimensions. The dashed line represents the baseline without bottleneck ($2.5 \times 10^{-2}$).}
    \label{fig:mse_sum}
\end{figure}
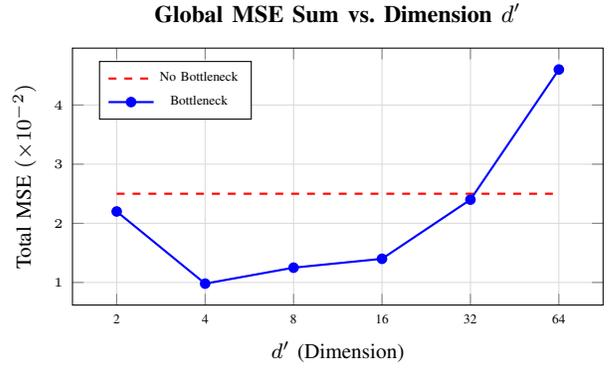

As illustrated in Figure~\ref{fig:mse_sum}, our model exhibits remarkable robustness to this constraint. Even with aggressive bottlenecks as small as $d'=4$ or $d'=8$, the model avoids the catastrophic failures typical of higher-dimensional autoregressive regimes. Interestingly, we observe that intermediate values of $d'$ can even lead to a reduction in long-horizon rollout error. This suggests that the bottleneck serves as a form of manifold regularization, effectively filtering out high-frequency numerical noise while preserving the essential low-dimensional physical dynamics $\mathcal{M}$.

\subsection{Temporal stability and spectral fidelity}
\label{sec:results_stability_spectral}

Pointwise rollout errors alone do not fully characterize system identification quality in turbulent regimes, where small phase shifts can inflate MSE while leaving the underlying dynamics largely intact. We therefore complement the target-choice ablation with two long-horizon diagnostics that directly probe autoregressive behavior: (i) temporal stability, quantified by frame-to-frame temporal-change magnitudes, and (ii) temporal spectral fidelity, computed from a fixed downstream probe signal (Section~\ref{sec:metrics}). We instantiate these diagnostics on \texttt{INC} and compare our best-performing target choice to ACDM under the same free-running rollout protocol. For fairness, we match ACDM’s training and sampling budget and use it as a reference diffusion baseline, not as the focus of a broader benchmarking study.

\begin{figure}[t]
  \centering
  \includegraphics[width=0.98\linewidth]{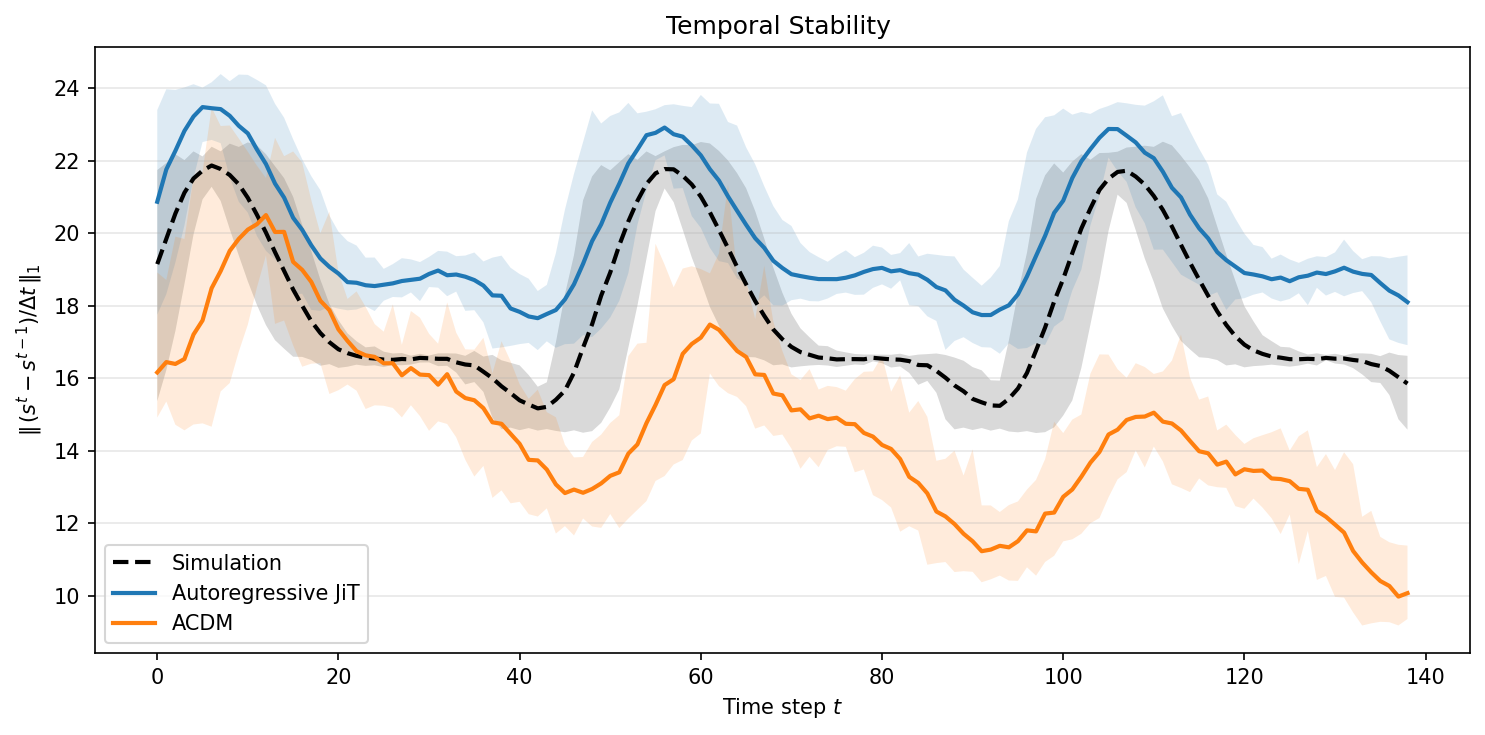}
  \vspace{-1mm}
  \caption{\textbf{Temporal stability under autoregressive feedback.}
    Magnitude of temporal change $\left\|\frac{\hat{x}_{t}-\hat{x}_{t-1}}{\Delta t}\right\|_1$ along free-running rollouts, compared to the reference simulation.}
  \label{fig:temporal_stability}
\end{figure}

\paragraph{Temporal stability.}
We assess stability via the per-step temporal-change magnitude $d_t$ (Eq.~\ref{eq:temporal_change}), i.e., the spatial mean of the absolute time increment $\left|\frac{\hat{x}_t-\hat{x}_{t-1}}{\Delta t}\right|$, aggregated along the rollout. Intuitively, this quantity tracks how ``active'' the generated dynamics remain under repeated self-conditioning: sustained increases indicate amplification of high-frequency temporal noise and drift, whereas sustained decreases indicate excessive damping and a progressive loss of dynamic content.
Figure~\ref{fig:temporal_stability} shows that our autoregressive rollouts follow the reference simulation closely across the horizon, including the characteristic oscillatory pattern associated with vortex shedding. ACDM exhibits a similar qualitative trend early on, but its temporal-change statistics deviate more noticeably as the rollout progresses, with a gradual drift away from the reference band toward the end of the horizon. Overall, the curves suggest that the proposed configuration maintains stable temporal activity over long autoregressive feedback, while ACDM is slightly more prone to long-horizon drift in this diagnostic.

\begin{figure}[t]
  \centering
  \includegraphics[width=0.98\linewidth]{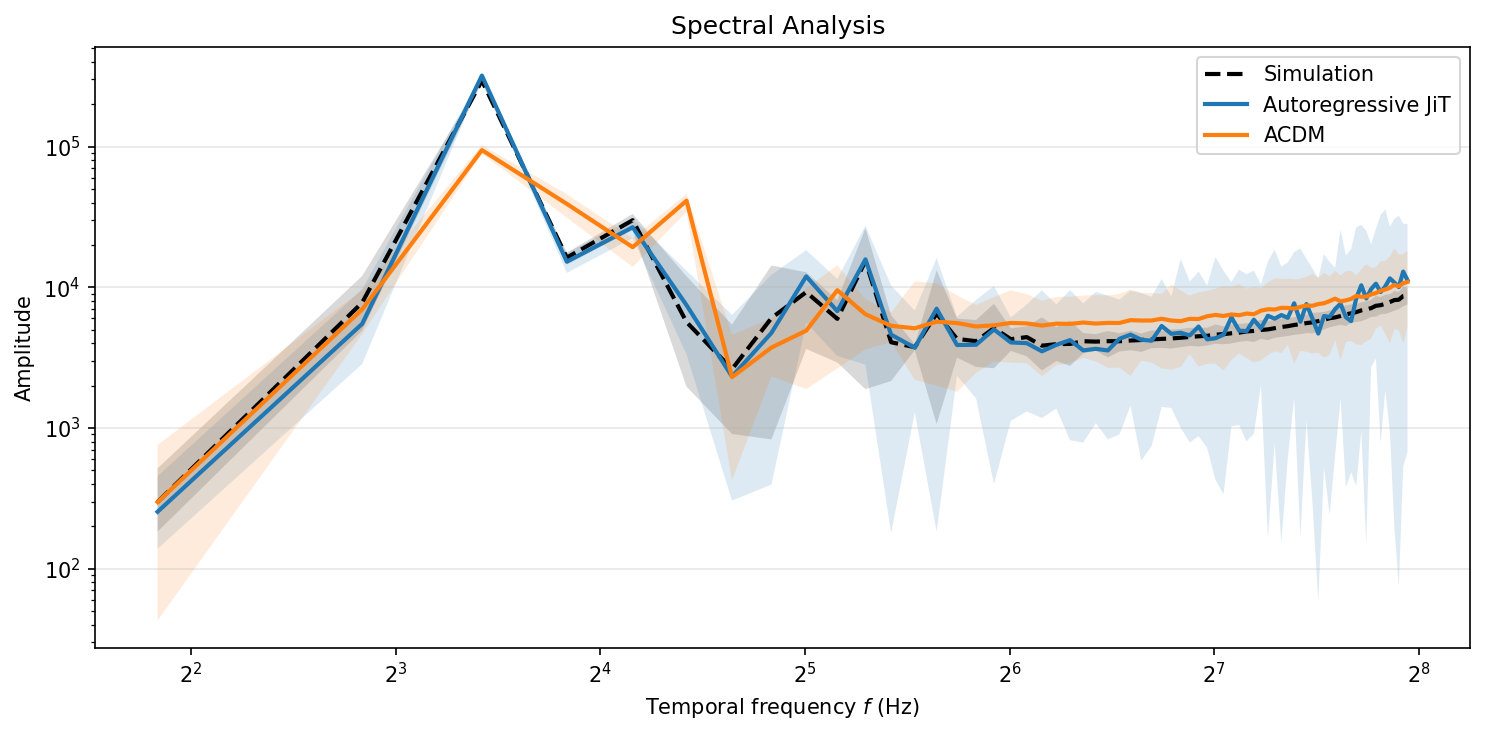}
  \vspace{-1mm}
  \caption{\textbf{Spectral fidelity of long rollouts.}
  Temporal frequency spectrum of the downstream probe signal (log scale), comparing generated rollouts to the reference simulation.}
  \label{fig:spectral_analysis}
\end{figure}

\paragraph{Spectral fidelity.}
To assess whether rollouts preserve the dominant temporal modes of the wake, we monitor a scalar probe signal extracted at a fixed downstream grid location $q=(i_q,j_q)$.
Concretely, we select a channel/component of interest at $q$ to form the time series $\{s_t\}_{t=0}^{H-1}$ (Eqs.~\ref{eq:probe_signal}--\ref{eq:probe_psd}) and compute its discrete Fourier spectrum.
Peaks in this spectrum correspond to coherent periodic content (vortex shedding and harmonics), while the higher-frequency tail reflects finer temporal fluctuations.
In Figure~\ref{fig:spectral_analysis}, the spectrum of our rollouts stays close to the reference across the dominant shedding band and much of the broadband region.
ACDM also captures the main low-frequency structure, but exhibits a more noticeable mismatch in parts of the spectrum, particularly at higher frequencies.
Together with the temporal-stability curves, this indicates that our autoregressive sampler better preserves the temporal organization of the wake over long horizons and maintains frequency statistics that remain consistent with the reference simulation under feedback.

\section{CONCLUSION}

We revisited target parameterization in diffusion-based identification of nonlinear systems with spatial outputs, using turbulent flow forecasting as a representative setting. Under matched backbones and training budgets, our experiments show that clean-state prediction yields more stable long-horizon rollouts and lower accumulated error than velocity and noise prediction. This advantage becomes more pronounced as per-token dimensionality increases, highlighting target parameterization as a key design choice for diffusion-based spatiotemporal modeling.

Several directions merit further study. First, it would be valuable to validate whether the observed advantage of clean-state prediction extends to broader classes of PDE-governed systems and experimental datasets. Second, future work should examine the interaction between target parameterization, patch size, and computational efficiency in settings where token count is not controlled. Finally, combining clean-state objectives with physics-informed constraints or uncertainty-aware identification could further improve robustness in challenging turbulent regimes.





\end{document}